\documentclass[11pt,a4paper]{article}

\usepackage{amssymb}
\usepackage{amsfonts}
\usepackage{amsmath}
\usepackage{latexsym,epsfig}
\usepackage{graphicx} 
\usepackage[usenames,dvipsnames]{color} 
\usepackage{hyperref}
\usepackage{breakurl} 
\usepackage{soul} 
\usepackage{array} 
\usepackage{verbatim}
\usepackage{colortbl} 
\usepackage[hang,small,bf]{caption} 
\usepackage{natbib}   
\usepackage[ruled,linesnumbered]{algorithm2e}  
\usepackage{url}
\usepackage{enumerate} 

\newcommand{\EndProof}{\hspace{\stretch{1}} $\Box$}
\newcommand\norm[1]{\left\lVert#1\right\rVert} 
\newcommand{\NP}{\mathcal{NP}}
\newcommand{\bigoh}{\mathcal{O}}
\newcommand{\LinkedIn}{\textsc{LinkedIn}}
\newcommand{\Flickr}{\textsc{Flickr}}
\newcommand{\Twitter}{\textsc{Twitter}}
\newcommand{\Answers}{\textsc{Answers}}
\newcommand{\Delicious}{\textsc{Delicious}}
\newcommand{\ResearchGate}{\textsc{ResearchGate}}
\newcommand{\Academia}{\textsc{Academia}}
\newcommand{\Facebook}{\textsc{Facebook}}
\newcommand{\Epa}{\textsc{Epa}}
\newcommand{\Rep}{\textsc{Rep}}
\newcommand{\Rip}{\textsc{Rip}}

\newtheorem{theorem}{Theorem}
\newtheorem{problem}{Problem}

\title{Synthetic Generation of Social Network Data \\ With Endorsements}

\author{Hebert P\'erez-Ros\'es$^{1, 2}$ and Francesc Seb\'e$^{1}$ \\ \\
$\phantom{0}^{1}$\emph{Dept. of Mathematics, Universitat de Lleida, Spain} \\
\vspace{2mm} 
$\phantom{0}^{2}$\emph{Conjoint Fellow, University of Newcastle, Australia} } 

\begin{document}
\maketitle

\vskip 5mm
\thispagestyle{empty}

\begin{abstract}

In many simulation studies involving networks there is the need to rely on a sample network to perform the simulation experiments. In many cases, real network data is not available due to privacy concerns. In that case we can recourse to synthetic data sets with similar properties to the real data. In this paper we discuss the problem of generating synthetic data sets for a certain kind of online social network, for simulation purposes. Some popular online social networks, such as \LinkedIn \ and \ResearchGate, allow user endorsements for specific skills. For each particular skill, the endorsements give rise to a directed subgraph of the corresponding network, where the nodes correspond to network members or users, and the arcs represent endorsement relations. Modelling these endorsement digraphs can be done by formulating an optimization problem, which is amenable to different heuristics. Our construction method consists of two stages: The first one simulates the growth of the network, and the second one solves the aforementioned optimization problem to construct the endorsements. 
\end{abstract}

\vskip 5mm

\section{Introduction}
\label{sec:intro}

Online social networks are ubiquitous in contemporary society. Ranging from general social networks, such as \Facebook \ or \Twitter, to the blogosphere and professional social networks, such as \LinkedIn, \ResearchGate, and \Academia, social networks now play a decisive role in politics, decision-making, marketing, mating, and information diffusion in general. 

An online social network can be effectively modelled by a graph, either directed or undirected, according to the nature of the  relationship established among its entities. For example, in \Facebook \ and \LinkedIn \ the nodes are the users' profiles, and the (symmetric) binary relation defined on the set of nodes is that of \lq friendship\rq \ or \lq acquaintance\rq. In this case, the ensuing graph is undirected, or symmetric. 

A second scenario is illustrated by the blogosphere, i.e the social network composed of blogs/bloggers and the (asymmetric) \lq recommendation\rq \ or \lq follower\rq \ relations among them, which gives rise to a directed graph. \ResearchGate \ is a professional social network, similar to \LinkedIn \ in many aspects, except in that the relation defined between two users is a \lq follower\rq \ relation, i.e. asymmetric. Analogous examples include \lq trust\rq \ statements in recommendation systems: some user states that he$/$she trusts the recommendations given by some other user. Additionally, weighted arcs appear in situations where such relations possess a certain degree of confidence (i.e. \lq trust\rq \ or \lq endorsement\rq \ statements could be partial). 

Some social networks, such as \LinkedIn \ and \ResearchGate, have recently included skills in users' profiles, and the possibility to endorse other users for a particular skill. For each particular skill, the endorsements make up a directed graph, which is a subgraph of the main graph of acquaintances. 

In many simulation studies involving social networks there is the need to count on a sample network to perform the simulation experiments \citep[e.g.][]{Coin07,Fow08,Fow09,Men08,Pan12,PSR,Sto02}. Due to privacy concerns, most social networks do not disclose sensitive information of its members to outsiders, which may include the set of acquaintances or the endorsements. Therefore, generating realistic synthetic networks stands out as an important challenge in social simulation. 

There are numerous models attempting to describe real-life networks. In particular, several models have been proposed to describe social networks. For instance, Leskovec \citep{Les08} suggests a model that describes quite accurately the dynamics of different online social networks, such as \Flickr, \Delicious, \Answers, and \LinkedIn. Leskovec's model is basically a simulation algorithm, which reproduces the arrival of new nodes, and the creation of new links among existing nodes, following a preferential attachment rule (see next section). 

However, Leskovec's model does not make any provision for endorsements, since that feature was added much later (2012 for \LinkedIn, and 2013 for \ResearchGate). As far as we know, there is no model of social networks that covers the endorsement feature. Including such an extension in Leskovec's model would require a comprehensive statistical study with real data over a relatively long period of time, just in the same manner that the model was created in the first place. Yet, it is possible to replace the aforementioned statistical analysis with a discrete optimization problem that uses a minimal amount of \emph{static} information. This optimization problem is amenable to different heuristics. 

In this paper we address the problem of generating realistic synthetic datasets that reproduce online social networks with the endorsement feature, such as \LinkedIn, for simulation purposes. We propose a method comprising two main steps: 

\begin{enumerate}[I]
\item The base graph, or graph of acquaintances, is created according to the network evolution model described in \citep{Les08}. 
\item Endorsements subdigraphs are generated by solving an optimization problem with the aid of heuristics. 
\end{enumerate}

For the basic terminology and notation of graphs, digraphs, and complex networks, we refer the reader to the Appendix. 


\section{Creating the network of acquaintances}
\label{sec:base}

In this section we briefly describe the network evolution algorithm given by \citep{Les08} for constructing the base network of acquaintances. 

We have rephrased the procedure described in \citep{Les08} as a discrete-event simulation algorithm (Algorithm \ref{algo:base}). 

\begin{algorithm}[ht]
\SetKwInOut{Input}{Input}
\SetKwInOut{Output}{Output}
\vspace{.2cm}
\Input{Termination conditions, node arrival function $A(t)$, and parameters $\alpha, \beta, \lambda$.} 
\Output{A base graph of acquaintances $G$.} 
\vspace{.2cm}
\tcc{ \,\,\, ----- \,\,\, INITIALIZATIONS \,\,\, -----}
Initialize $G$\;
Initialize priority queue $Q$\;
\ForEach{vertex $v \in G$}{
Generate random lifetime and sleeping time for $v$\;
Push $v$ onto $Q$\;
}
Initialize clock\;
\tcc{ \,\,\, ----- \,\,\, MAIN CYCLE \,\,\, -----}
\While{ termination conditions not met }{
  Pop a vertex $u$ from $Q$\;
  Set clock to $u$'s wake-up time\;
  \If{clock does not exceed lifetime of $u$}{
       Create random two-hop edge starting at $u$\;
       Generate random sleeping time for $u$\;
       Push $u$ back onto $Q$\;
   }
  Create set of vertices $S$ that have appeared in the meantime\;
  \ForEach{vertex $v \in S$}{
      Link $v$ to some $w \in G$ with preferential attachment\;
      Generate random lifetime and sleeping time for $v$\;
      Push $v$ onto $Q$\;
   }
}

\caption{Base network generation}
\label{algo:base}  
\end{algorithm}

The algorithm receives as input the termination conditions, and a set of network-dependent parameters. The values of the parameters determined empirically in \citep{Les08} for different social networks are given in Table \ref{tab:parameters}.

\begin{table}[htp]
\begin{center}
\begin{tabular}{|cc|*{4}{c}|} \hline
\multicolumn{2}{|c|}{Network} & $A(t)$ & $\alpha$ & $\beta$ & $\lambda$\\ \hline
\rowcolor[gray]{.9} 
\multicolumn{2}{|c|}{\Flickr} & $e^{0.25 t}$ & 0.84 & 0.002 & 0.0092\\ 
\multicolumn{2}{|c|}{\Delicious} & $16t^2+3000t+40000$ & 0.92 & 0.00032 & 0.0052\\ 
\rowcolor[gray]{.9} 
\multicolumn{2}{|c|}{\Answers} & $-4544t^2+160000t-2500$ & 0.85 & 0.0038 & 0.0019\\ 
\multicolumn{2}{|c|}{\LinkedIn} & $3900t^2+76000t-130000$ & 0.78 & 0.00036 & 0.0018\\ \hline
\end{tabular}
\caption{Parameters of different networks. Node arrivals are measured monthly; the other parameters are adjusted for days.}
\label{tab:parameters}
\end{center}
\end{table}

Lines 1 to 7 make up the initialization block. The graph $G$ may be initialized as a small clique, which is quite a realistic assumption. In our case, we start with a small complete graph of five nodes. 

Lifetimes (in days) are randomly generated from the exponential distribution $f(x) = \lambda e^{\lambda x}$. Sleeptimes (in days) are randomly generated from the exponential distribution $g_{d, \alpha, \beta}(x) = \frac{1}{C_{d, \alpha, \beta}} x^{-\alpha} e^{-\beta d x}$, where $d$ is the degree of the node in question, and $C_{d, \alpha, \beta}$ is the normalizing constant. 

Next comes the main cycle. Possible termination conditions for this cycle are: 
\begin{enumerate}
\item the clock exceeds a certain pre-defined simulation time, 
\setlength{\itemsep}{0pt}
\item a pre-defined number of iterations is reached,  
\setlength{\itemsep}{0pt}
\item the network reaches a pre-defined number of nodes,
\end{enumerate}
or any combination of them.

\begin{figure}[htbp]
 \begin{center}
  \includegraphics[width=1.0\textwidth]{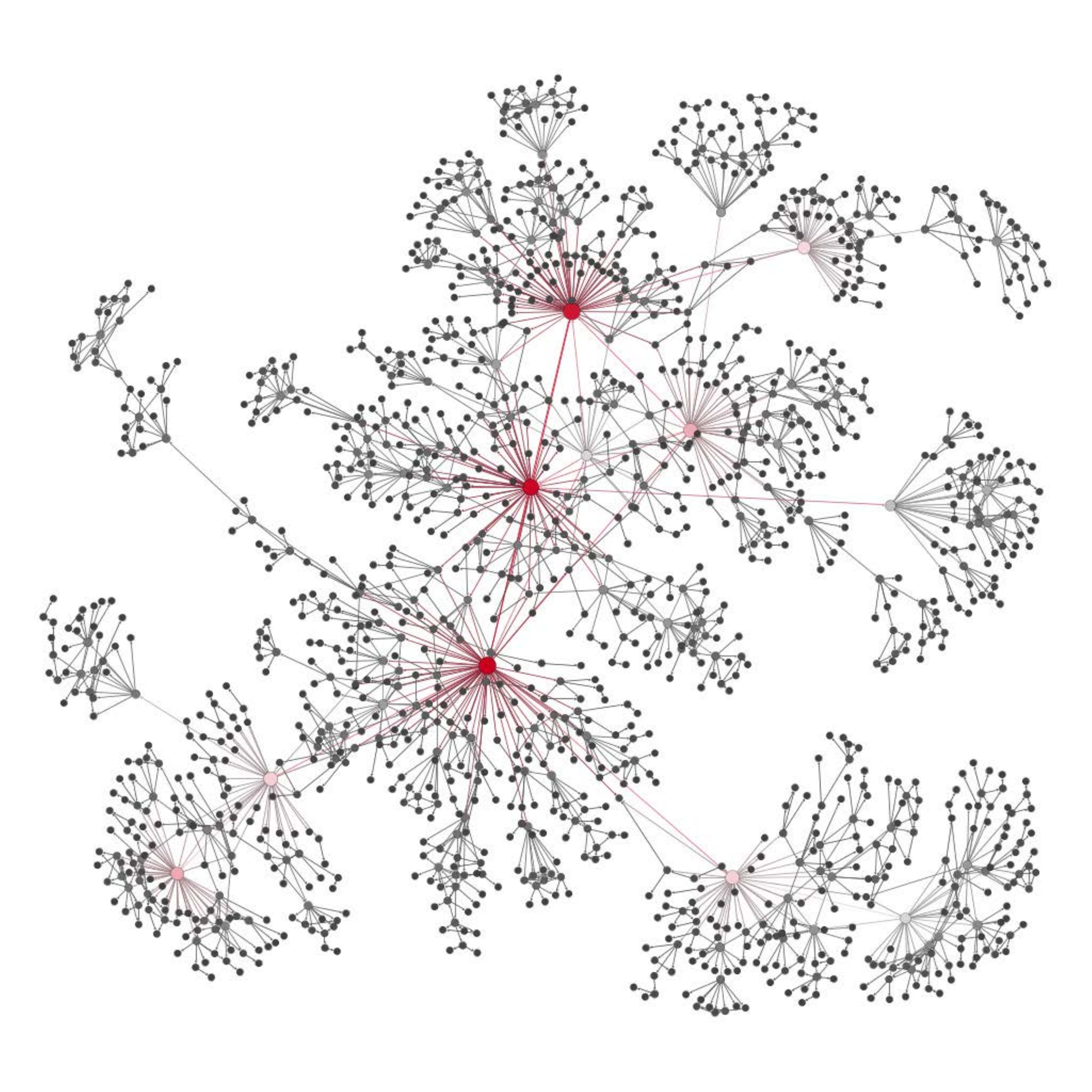}
 	 \caption{Sample network of 1493 nodes and 2489 links generated by Algorithm \ref{algo:base}. The graph $G$ was initialized as a clique of five nodes.}
 	 \label{fig:net45}
 \end{center}	
\end{figure}

The networks that arise from this process are scale-free, with power-law degree distribution with exponent $1+ \frac{\lambda \Gamma(2-\alpha)}{\beta \Gamma(1-\alpha)}$, where $\Gamma$ is the Gamma function. The density of these networks (i.e. their average degree) increases over time, while their diameter actually \emph{decreases}. Figure \ref{fig:net45} displays a network obtained by $1000$ iterations of the main cycle in Algorithm \ref{algo:base}. 

\section{Generating the endorsement subdigraphs}
\label{sec:endorsements}

As it was mentioned above, the model given in \citep{Les08} does not consider endorsements, and no other model does, to the best of our knowledge. In principle it might be possible to study the emergence and evolution of endorsements, in much the same way that \citep{Les08} studies the dynamics of their underlying graph of acquaintances. This, however, appears like a formidable task. To begin with, it requires access to the real data, from the moment that the endorsement feature was introduced. 

On the other hand, it is not too difficult to take a \emph{static snapshot} of the endorsements at a given time. Let us assume that we are interested in a set $S$ of skills ($\vert S \vert = n_s$). From now on $G$ will denote the real social network at a given time, while $G'$ will be the synthetic base graph created by Algorithm \ref{algo:base}. With the aid of any statistical sampling method we can estimate the frequencies and relative co-occurrences of the skills $S$ in $G$. Then it is possible to reproduce these statistics in synthetically generated subgraphs of $G'$. 

Network sampling methods have been investigated since at least the 1960s \citep[e.g.][]{Frank77,Frank78,Good61}, and several effective sampling methods have been devised, such as \emph{snowball} (a variant of breadth-first search) \citep{Frank77,Good61}, \emph{random walk} \citep{Har13,Kur11,Lu12,Rib10}, or \emph{forest fire} \citep{Les05,Les06}. The interested reader can find up-to-date discussions in \citep{Doe13,Oka11,Wang11}. At any rate, some (rather informal) experiments suggest that an accurate estimation of network statistics requires a sample of size at least 15\% of the total network size \citep{Frank12}, which is beyond our current capabilities in the case of \LinkedIn. Nevertheless, our main concern at this point is not an accurate sampling of the network, but \emph{an accurate replication of the sampling results}. Therefore, in our experiments we have opted for taking a small sample of a few hundred nodes by a simple variant of breadth-first sampling, and we make no claims of accuracy in that respect. 

All the relevant information relative to endorsements can be encoded into an $n_s \times n_s$ matrix $\mathbf{M}_S(G) = (m_{ij})$. If the set of skills $S$ is fixed, we can drop the subscript $S$ for simplicity of notation. The diagonal entry $m_{ii}$ represents the ratio between the number of nodes that are endorsed for the $i$-th skill and the total number of nodes in the graph. On the other hand, the entry $m_{ij}$, for $i \neq j$, is the ratio between the number of nodes that have been endorsed for both skills, $i$ and $j$, and the number of nodes that have been endorsed for the $i$-th skill alone. Thus, $m_{ij}$ may be taken as an estimate of the conditional probability that a user gets endorsed for skill $j$, given that it is endorsed for skill $i$.

Now our problem can be formulated as follows: 

\begin{problem}[Replication of endorsement pattern]--
\label{prob:endor}
Given a finite un-directed graph $G' = (V', E')$, a set of skills $S$, with $\vert S \vert = n_s$, and a matrix $\mathbf{M} = (m_{ij})_{n_s \times n_s}$ with non-negative rational coefficients, determine if there exists a configuration of endorsements such that $\mathbf{M}(G') = \mathbf{M}$. In the affirmative case, we say that $\mathbf{M}$ is \emph{realizable}, or that it is an \emph{endorsement pattern}. 
\end{problem}

Problem \ref{prob:endor} (or \Rep, for short) turns out to be closely related to the problem of \emph{recognizing intersection patterns of sets} \citep{Ch80}. In \emph{Intersection Pattern Recognition} (\Rip) we also have a (symmetric integral) matrix $\mathbf{P} = (p_{ij})_{n \times n}$, where the $p_{ij}$ are non-negative integers, and the goal is to find a collection of $n$ arbitrary sets, $S_1, S_2, \cdots S_n$, such that $\vert S_i \cap S_j \vert = p_{ij}$ for all $1 \leq i,j \leq n$. Obviously, the diagonal elements $p_{ii}$ correspond to the set cardinalities $\vert S_i \vert$. If that collection exists, then the matrix $\mathbf{P}$ is called an \emph{intersection pattern}, or in analogy with the terminology of Problem \ref{prob:endor}, we also say that $\mathbf{P}$ is \emph{realizable}. The collection $S_1, S_2, \cdots S_n$ is called a \emph{realization}. 

Deciding whether a given matrix $\mathbf{P}$ is an intersection pattern is $\mathcal{NP}$-complete unless $p_{ii} \leq 2$ for all $1 \leq i \leq n$. Recall that $\NP$-complete problems are computationally intractable with today's technology. Since there is a straightforward polynomial-time reduction from \Rip \ to \Rep, we are led to the following result:  

\begin{theorem}
\label{theo:complete}
-- \Rep \ is $\NP$-complete. 
\end{theorem}

\noindent \textbf{PROOF.} It is a routine task to verify that \Rep \ $\in \NP$ for any finite $G'$. Hence we only have to check that \Rep \ is $\NP$-hard by exhibiting a polynomial-time reduction from \Rip \ to \Rep. Given an arbitrary symmetric integral matrix $\mathbf{P}$, we must produce an undirected graph $G'$, and a matrix $\mathbf{M} = (m_{ij})_{n_s \times n_s}$ with non-negative rational coefficients,  so that $\mathbf{P}$ is realizable (as an intersection pattern) if, and only if, $\mathbf{M}$ is realizable (as an endorsement pattern on $G'$). 

Since the diagonal elements of $\mathbf{P}$ correspond to the cardinalities of the sets $S_i$, we can say without any loss of generality that the trace of $\mathbf{P}$ is an upper bound for $\vert \mathcal{U} \vert$, where $\mathcal{U} = \bigcup_{i=1}^n S_i$. In other words, a given matrix $\mathbf{P}$ is an intersection pattern if, and only if, there exists a realization with size less than or equal to $\mbox{tr}(\mathbf{P})$. 


Thus, let $G'$ be a connected graph on $\mbox{tr}(\mathbf{P})$. In \Rip \ we will choose the sets $S_1, S_2, \ldots, S_n$ among the vertices of $G'$. We make $n_s$ equal to $n$, and $m_{ij} = p_{ij}/p_{ii}$ for all $1 \leq i,j \leq n$ such that $i \neq j$. Finally, make $m_{ii} = p_{ii}/\vert V' \vert$ for all $1 \leq i \leq n$. Obviously, $\mathbf{M}$ is an endorsement pattern over $G'$ if, and only if, $\mathbf{P}$ is an intersection pattern. This completes the polynomial-time reduction. \EndProof
\\\\
Problem \ref{prob:endor} (a decision problem) can also be formulated as a combinatorial optimization problem. Let $\delta$ be a fixed given matrix \lq distance\rq \ function. The approximation$/$optimization version of \Rep \ is called Endorsement Pattern Approximation (\Epa):

\begin{problem}[Endorsement pattern approximation]--
\label{prob:approx}
Given a finite un-directed graph $G' = (V', E')$, a set of skills $S$, with $\vert S \vert = n_s$, and a matrix $\mathbf{M} = (m_{ij})_{n_s \times n_s}$ with non-negative rational coefficients, find a configuration of endorsements that minimizes $\delta(\mathbf{M}, \mathbf{M}')$, where $\mathbf{M}' = \mathbf{M}(G')$.    
\end{problem}

In our experiments we have considered a family of distance functions $\delta_\mathbf{W}$, parametrized by a positive matrix $\mathbf{W}$, defined as 

\begin{equation}
\label{eq:distancefun}
\delta_\mathbf{W}(\mathbf{M}, \mathbf{M}') = \frac{1}{n_s^2} \norm{ \mathbf{W} \circ (\mathbf{M}-\mathbf{M}') },
\end{equation}


where $\norm{ \cdot }$ is the squared Frobenius matrix norm metric:

\begin{displaymath}
\norm{A} = \sum_{i=1}^n \sum_{j=1}^n |a_{ij}|^2 = \mbox{tr}(A^TA)
\end{displaymath}

\citep[see e.g.][p. 217]{deza13}, and \lq$\circ$\rq \ denotes Hadamard (or elementwise) matrix multiplication. 

Equation \ref{eq:distancefun} is merely a weighted normalized error of the variable matrix $\mathbf{M}'$ with respect to the fixed \emph{target matrix} $\mathbf{M}$. The purpose of the weight matrix $\mathbf{W}$ is to control the relative importance of the different pieces of information encoded into $\mathbf{M}(G)$. In particular, we have chosen $\mathbf{W}$ so as to confer more importance to the diagonal elements of $\mathbf{M}'$, representing the individual frequences of the skills within $G'$. 

Theorem \ref{theo:complete} confirms that reproducing endorsements is computationally hard, either in its exact form or in the approximate form, which justifies the use of heuristics to address the problem. In the next section we describe some simple heuristic methods that have shown to produce good results. 

\section{Computational experiments}
\label{sec:experiments}

In order to test the above ideas in practice we have focused on \LinkedIn, which to the best of our knowledge, was the first professional social network to introduce the endorsement feature. So, let $G$ be the real \LinkedIn \ network as of Sep. 15, 2013. We have implemented Algorithm \ref{algo:base} and used it to generate an undirected network of contacts $G'$ with $1 493$ nodes and $2 489$ edges. This network is small enough to be tractable, and yet large enough to derive meaningful conclusions. 

\subsection{An example}
\label{sec:experiment1}

We first illustrate our experimental design with a small example, taken from \citep{PSR}. The example considers only five skills: 1. Programming, 2. C++, 3. Java, 4. Mathematical Modelling, 5. Statistics. Those skills were chosen in \citep{PSR} for two main reasons: 
\begin{enumerate}
\item Those five skills abound in the \LinkedIn \ sample taken. 
\item Those five skills can be clearly separated into two groups or clusters, namely programming-related skills, and mathematical skills, with a large intra-cluster correlation, and a smaller inter-cluster correlation. This is a small-scale representation of the real network, where skills can be grouped into clusters of related skills, which may give rise to different patterns of interaction among skills. 
\end{enumerate}

The occurrences and co-occurrences of the five skills were computed in \citep{PSR} for a small community of \LinkedIn \, members, resulting in the matrix $\mathbf{M}$ of Eq. \ref{eq:obvserved1}.  

\begin{equation}
\label{eq:obvserved1}
\mathbf{M} = \left( \begin{array}{ccccc}
0.12 & 0.42 & 0.42 & 0.5  & 0.33 \\
0.62 & 0.08 & 0.62 & 0.25 & 0.12 \\
0.62 & 0.62 & 0.08 & 0.12 & 0.12 \\
0.75 & 0.25 & 0.12 & 0.08 & 0.5 \\
0.5  & 0.12 & 0.12 & 0.5  & 0.08 
\end{array} \right)
\end{equation} 

We are now done with \LinkedIn; from now on we are going to work on the synthetic base network generated by Algorithm \ref{algo:base}. For each skill we want to construct a random endorsement digraph (a random sub-digraph of the base network), in such a way that the above frequencies are respected. This can be done efficiently by means of a simple local search heuristic. 

The local search algorithm starts with a set of random endorsement subdigraphs of $G$, and then refines the endorsements by small local changes, thus improving the approximation to $\mathbf{M}$ at each step. The local search step consists either in the random addition or deletion of an arc in some endorsement digraph chosen at random. This algorithm stops either when some approximation threshold is reached, or when it is not possible to improve (i.e. decrease) the objective function after a certain number of trials (500 in our case). Algorithm \ref{algo:localsearch} formalizes these ideas.  

\begin{algorithm}[htp]
\SetKwInOut{Input}{Input}
\SetKwInOut{Output}{Output}
\vspace{.2cm}
\Input{Base network $G = (V, E)$, an $n_s \times n_s$ matrix $\mathbf{M}$, and termination conditions.} 
\Output{A set of $n_s$ endorsement subdigraphs $D = D_1, \ldots, D_{n_s}$ of $G$} 
\vspace{.2cm}
\tcc{ \,\,\, ----- \,\,\, INITIALIZATIONS \,\,\, -----}
Initialize $D_1, \ldots, D_{n_s}$ at random\;
Compute the matrix $\mathbf{M}' = \mathbf{M}(G)$ from $G$ and $D_1, \ldots, D_{n_s}$\;
\tcc{ \,\,\,\,\,\,\, ----- \,\,\, MAIN CYCLE \,\,\, -----}
\While{ termination conditions not met }{
  Choose $i$ at random, with $1 \leq i \leq n_s$\;
  Choose one of two actions at random: \textsc{Insert} or \textsc{Delete}\;
  \eIf{\textsc{Insert}}{
       Choose a random edge $(u,v) \in E$ such that there is no arc $u \rightarrow v$ in $D_i$\;
       Compute the matrix $\mathbf{M}''$ after the addition of $u \rightarrow v$ to $D_i$\;
       \If{ $\delta(\mathbf{M}'',\mathbf{M}) < \delta(\mathbf{M}',\mathbf{M})$ }{
             Insert $u \rightarrow v$ in $D_i$\;
             $\mathbf{M}' := \mathbf{M}''$\;
             }
    }{        
       Choose a random arc $u \rightarrow v$ in $D_i$\;
       Compute the matrix $\mathbf{M}''$ after the deletion of $u \rightarrow v$ from $D_i$\;
       \If{$\delta(\mathbf{M}'',\mathbf{M}) < \delta(\mathbf{M}',\mathbf{M})$}{
             Delete $u \rightarrow v$ from $D_i$\;
             $\mathbf{M}' := \mathbf{M}''$\; 
             }      
     }
}

\caption{Local search algorithm for creating the endorsement digraphs}
\label{algo:localsearch}  
\end{algorithm}

With the aid of Algorithm \ref{algo:localsearch}, the matrix given in Eq. \ref{eq:obvserved1} was approximated to within an error of $10^{-5}$. The base network generated by Algorithm \ref{algo:base}, together with the endorsement digraphs obtained by Algorithm \ref{algo:localsearch} for this example can be downloaded from \url{http://www.cig.udl.cat/sitemedia/files/MiniLinkedIn.zip}. 

\subsection{Further experimental results}
\label{sec:experiment2}

Now, the purpose of our computational experiments is to check that our heuristic method above scales up to larger instances of the problem. We keep the same base network, but we introduce a larger number of skills (and consequently, a larger endorsement pattern matrix). There are two main risks threatening the convergence of Algorithm \ref{algo:localsearch}: 

\begin{enumerate}
\item The pattern matrix $\mathbf{M}$ is not realizable, and 
\item The local search procedure gets trapped in a local minimum of the objective function.  
\end{enumerate}

At any rate, if the pattern matrix $\mathbf{M}$ arises from actual network statistics, such as in the previous example, it should be approximable to within a reasonable threshold. The second situation described above arises in any heuristic algorithm attempting to solve an $\NP$-hard problem, and many strategies have been devised over the years to minimize that risk. We cannot cover all those strategies in depth here, but we do mention one possible solution at the end of this section. The interested reader can find a wealth of information in \citep{Aarts97,metahandbook} and other sources. 

Thus, for our experiments we have fixed a number of skills $n_s$, and we have generated $n_s$ random endorsement subdigraphs of the synthetic base graph $G'$, and then we have computed the corresponding pattern matrix $\mathbf{M}(G')$ (which we know that is realizable, since it arises from actual endorsement digraphs). Then we applied Algorithm \ref{algo:localsearch} on $\mathbf{M}(G')$, and measured its performance. 

The initial endorsement subdigraphs of $G$ were generated as follows: For each endorsement digraph $D_i$, edges were picked at random from $G'$, were assigned a random orientation and were added to $D_i$, until the number of vertices with positive in-degree in $D_i$ was close to $m_{ii}$. The cost of this initialization is $\bigoh(\vert E \vert)$. 

We have experimented with different values of $n_s$ ($n_s$ = 5, 10, 15, 20, 30, 40, 50) and different probability distributions for the arcs of the endorsement digraphs, namely with uniform probabilities in the ranges $(0,1)$,  $(0,0.35)$, $(0.35, 0.65)$, $(0.65, 1)$. 

\begin{figure}[htbp]
 \begin{center}
  	\includegraphics[width=1.0\textwidth]{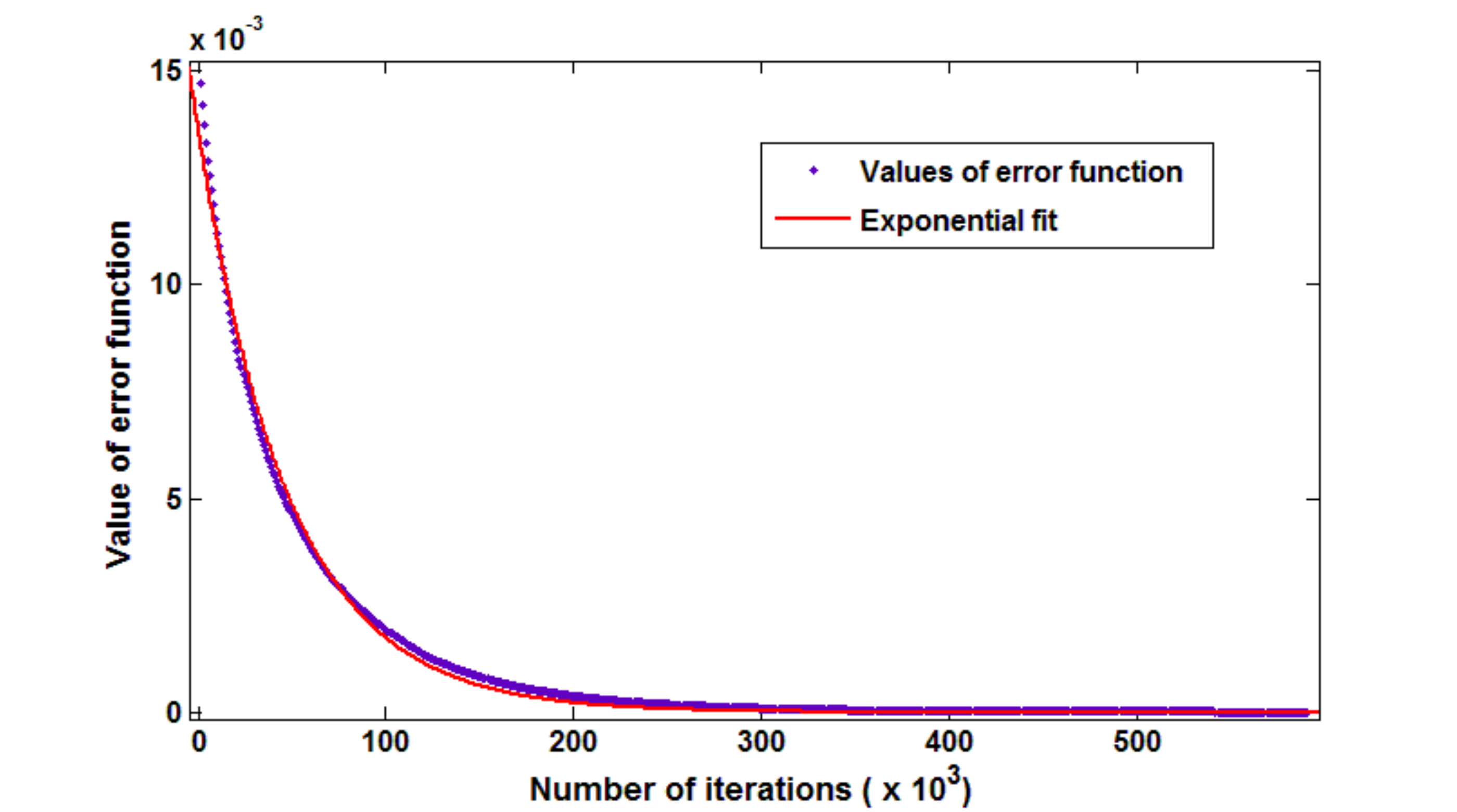}
 	 \caption{Typical evolution of the error function}
 	 \label{fig:fit}
 \end{center}	
 \end{figure}

Algorithm \ref{algo:localsearch} converged in all cases with an exponential convergence rate, i.e. the distance function decreased according to the rule $y = ae^{bx}$, with $a>0$ and $b<0$. Figure \ref{fig:fit} shows the convergence rate for one particular experiment, namely $n_s = 50$, and all frequencies taken uniformly in the range $(0.35, 0.65)$. As it can be seen, the values of the distance function can be fitted almost exactly by an exponential $y = ae^{bx}$, with $a = 0.01364$ and $b = -0.02055$. 

\begin{figure}[htbp]
 \begin{center}
  	\includegraphics[width=1.0\textwidth]{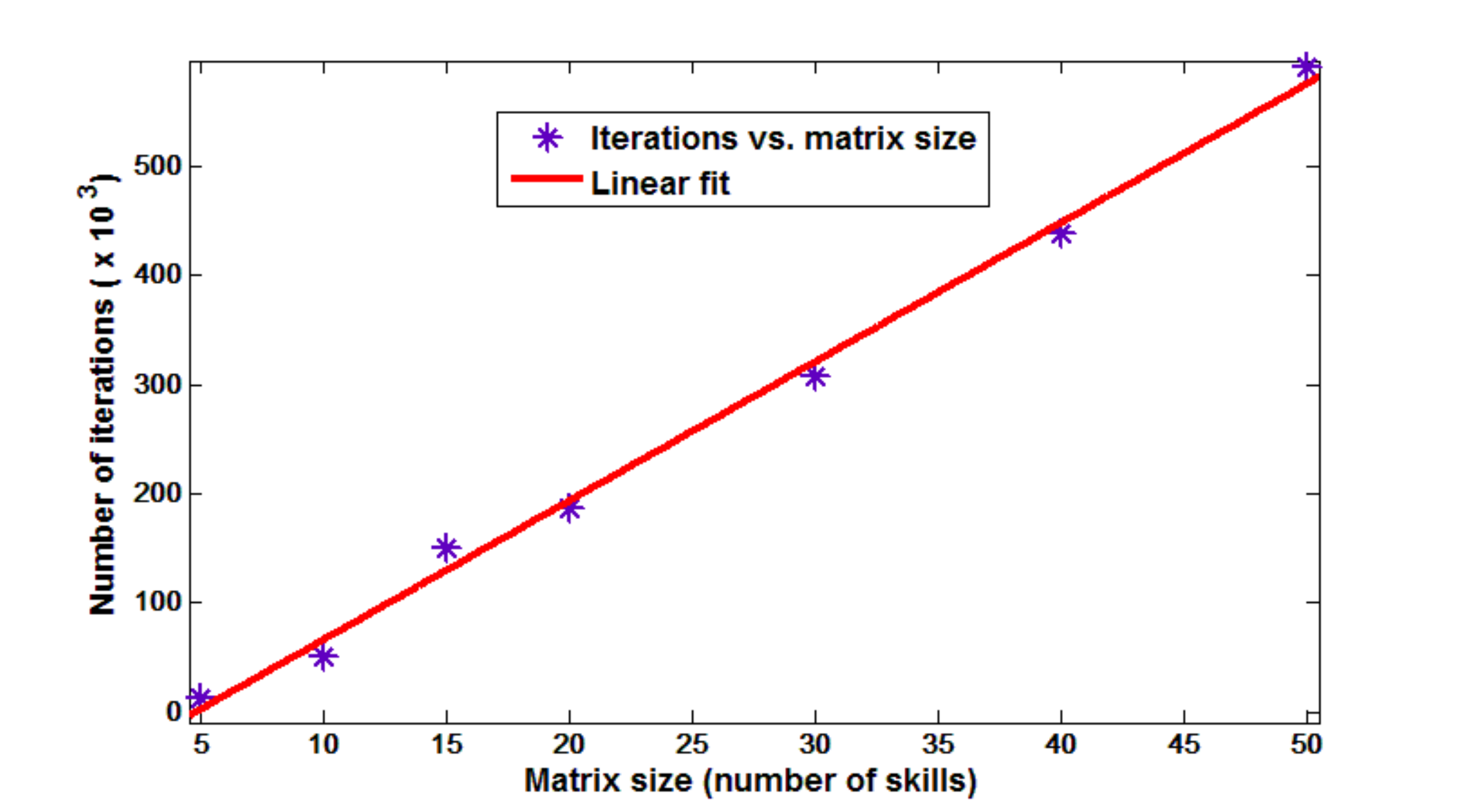}
 	 \caption{Number of iterations as a function of $n_s$}
 	 \label{fig:runningtime}
 \end{center}	
 \end{figure}

With such a fast convergence rate, the running time of the algorithm behaves as a linear function of $n_s$. Figure \ref{fig:runningtime} shows the behaviour of the running time (or more precisely, the number of iterations) as a function of the matrix size. The values correspond to the mean number of iterations for random matrices with values in the range $(0.35, 0.65)$. The linear regression fit yields the function $y = 12.76 x - 61.95$, which represents the data with high accuracy. 

The computational results obtained above provide criteria to fine-tune the termination conditions of Algorithm \ref{algo:localsearch}. For example, they suggest that a threshold of $10^{-5}$ is a reasonable criterion for termination. 

In order to reduce the risk of getting trapped in a local minimum, we can initialize the endorsement digraphs (Step 1 of Algorithm \ref{algo:localsearch}) in such a way that the matrix $\mathbf{M}'$ is as close as possible to $\mathbf{M}$. There is of course a tradeoff between the approximation of $\mathbf{M}'$ to $\mathbf{M}$ and the computational effort involved, but there are some simple heuristics that can give us a reasonably close initial approximation to $\mathbf{M}(G)$ with little computational effort. 

For example, we may construct a better first approximation to $\mathbf{M}$ by a random \lq greedy\rq \ procedure: Pick an edge at random, assign an orientation to it, and assign the resulting arc to some endorsement digraph, so that the assignment minimizes the distance function given in Eq. \ref{eq:distancefun2}. 

\begin{equation}
\label{eq:distancefun2}
\rho_\mathbf{W}(\mathbf{M}, \mathbf{M}') = \norm {\mathbf{W} \circ (\mathbf{M}-\mathbf{M}')},
\end{equation}

where again $\norm{ \cdot }$ denotes the squared Frobenius matrix norm metric, and \lq$\circ$\rq \ denotes Hadamard matrix multiplication. 

This procedure stops either when some approximation threshold is reached, or when the distance function starts to increase steadily. The cost of this initialization is slightly higher than the previous one, but asymptotically it remains $\bigoh(\vert E \vert)$. 

\subsection{Running time and scalability}
\label{sec:scalability}

In order to assess the actual running times, and thus appraise the feasibility and scalability of our algorithms in practice, we have performed an additional set of experiments. Algorithm \ref{algo:base} was run with a number of iterations ranging from $250$ to $2000$, at intervals of $250$. For each number of iterations, the algorithm was run five times, thus getting five different base networks. Among these five networks we have chosen the one with the median number of vertices, and have used it as input for Algorithm \ref{algo:localsearch}, so as to attach endorsements to it. 

The experiments were performed on very affordable hardware, namely an Acer Aspire 5749 laptop computer, equipped with an Intel Core i3-2330M processor, and 4 GB RAM. The statistics of the experiments are collected in Table \ref{tab:actualtimes}. 

\begin{table}[htp]
\begin{center}
\scalebox{1.0}{ 
\begin{tabular}{|l|llllllll|} \hline
Num. iterations & 250 & 500 & 750 & 1000 & 1250 & 1500 & 1750 & 2000 \\  
\rowcolor[gray]{.9}
Num. nodes & 664 & 925 & 1185 & 1364 & 1605 & 1644 & 1824 & 1925 \\ 
Num. links & 918 & 1430 & 1932 & 2361 & 2852 & 3147 & 3567 & 3919 \\ 
\rowcolor[gray]{.9} 
Time 1 & 20 & 38 & 61 & 93 & 131 & 137 & 181 & 200 \\ 
Time 2 & 8 & 22 & 44 & 69 & 108 & 126 & 164 & 211 \\ 
\rowcolor[gray]{.9}
Total time & 28 & 60 & 105 & 162 & 239 & 263 & 345 & 411 \\ \hline 
\end{tabular} } 
\caption{Network generation times. \lq Time 1\rq \ is the time taken to generate the base network (in seconds). \lq Time 2\rq \ is the time taken to add the endorsements to the base network (in seconds). The total time is just the sum of \lq Time 1\rq \ and \lq Time 2\rq; it does not include other subsidiary tasks, such as drawing the network.}
\label{tab:actualtimes}
\end{center}
\end{table}

For many practical purposes, a network of $1925$ nodes is large enough, and it can be generated and fitted with endorsements in about $7$ minutes. If we wanted a larger network we just have to run the algorithm for a longer time. The base network, together with the endorsements, can be stored in less than $30$ KB using a data structure based on adjacency lists. 

\begin{figure}[htbp]
 \begin{center}
  	\includegraphics[width=1.0\textwidth]{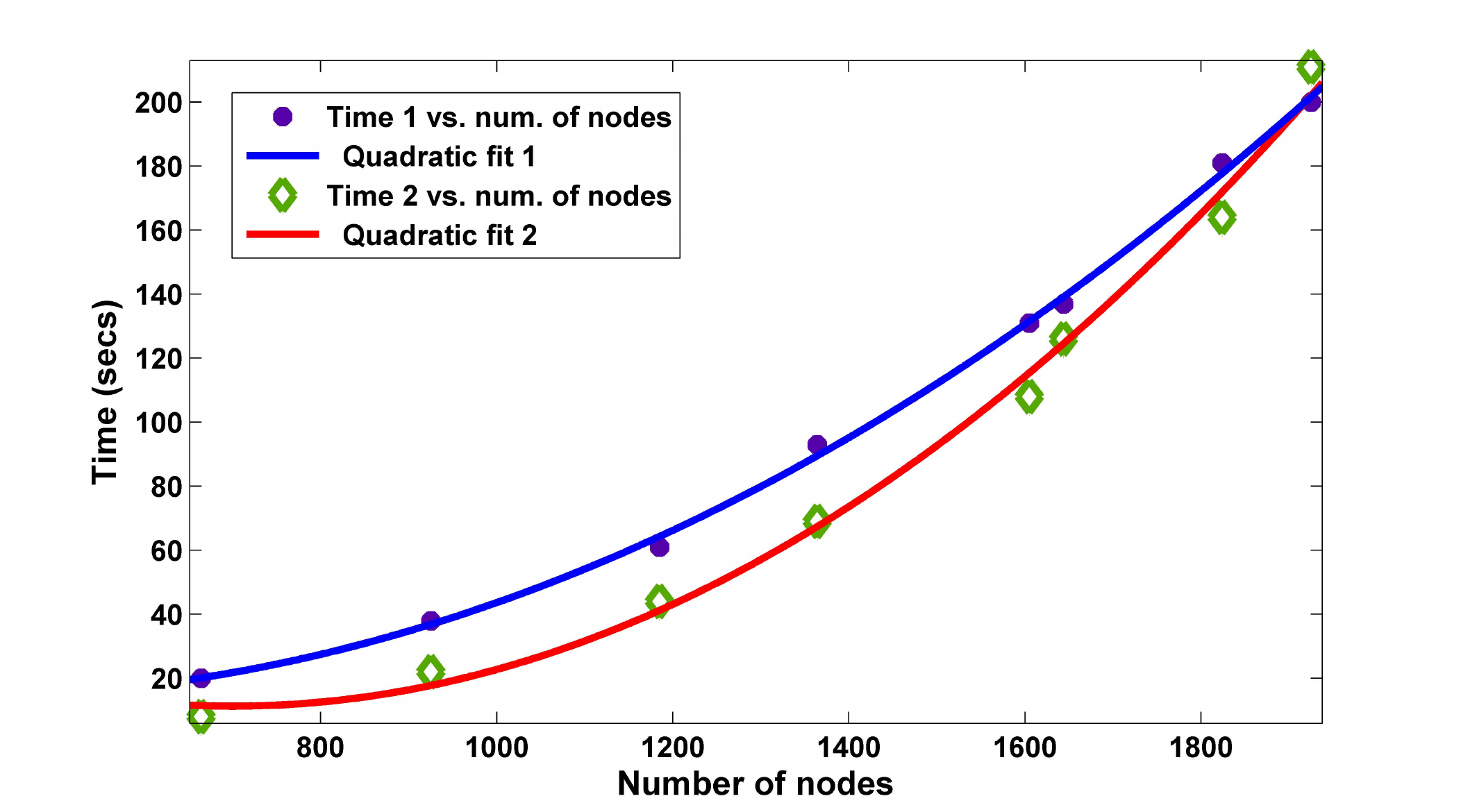}
 	 \caption{Running time as a function of the number of nodes. \lq Time 1\rq \ is the running time of Algorithm \ref{algo:base}. \lq Time 2\rq \ is the running time of Algorithm \ref{algo:localsearch}.}
 	 \label{fig:actualtimes}
 \end{center}	
 \end{figure}

On the basis of the data collected in Table \ref{tab:actualtimes} we conclude that the running time of Algorithm \ref{algo:base} is best approximated by the quadratic function $0.0000799 \times N^2 - 0.06303 \times N + 26.75$, where $N$ is the number of vertices of the network produced, while the running time of Algorithm \ref{algo:localsearch} follows a function $0.00008451 \times N^2 - 0.1778 \times N + 73.5$. Figure \ref{fig:actualtimes} provides a visual comparison between the two running times. 

By extrapolating these functions we can formulate predictions on the actual time that it will take to generate a network of some specified size, assuming that the behaviour of the programs remains stable. For example, we should be able to generate a network of about $10,000$ nodes in five hours approximately (on the same hardware), and a network of about $100,000$ nodes in a bit more than two weeks, provided that we had enough memory to store it. Hence, the practical limit of our implementation would be somewhere in the order of $X \times 10^4$ nodes, which is satisfactory for a large number of practical situations. Obviously, this limit can be greatly expanded if we migrate to a more powerful computer. 

\section{Conclusions and future research}
\label{sec:open}

In this paper we have shown how to synthesize a network similar to the social network \LinkedIn, for simulation purposes. The construction process consists of two stages: 

\begin{enumerate}[I]
\item Construction of the base network, and 
\item Addition of the endorsements.  
\end{enumerate}

The first stage is formulated and implemented as a discrete-event simulation algorithm, which is interesting in its own right. Nevertheless, our main contribution resides in the second stage, where the addition of endorsements is formulated in terms of a discrete optimization problem, which is then efficiently solved via heuristics. 

\begin{figure}[htbp]
 \begin{center}
  	\includegraphics[width=0.7\textwidth]{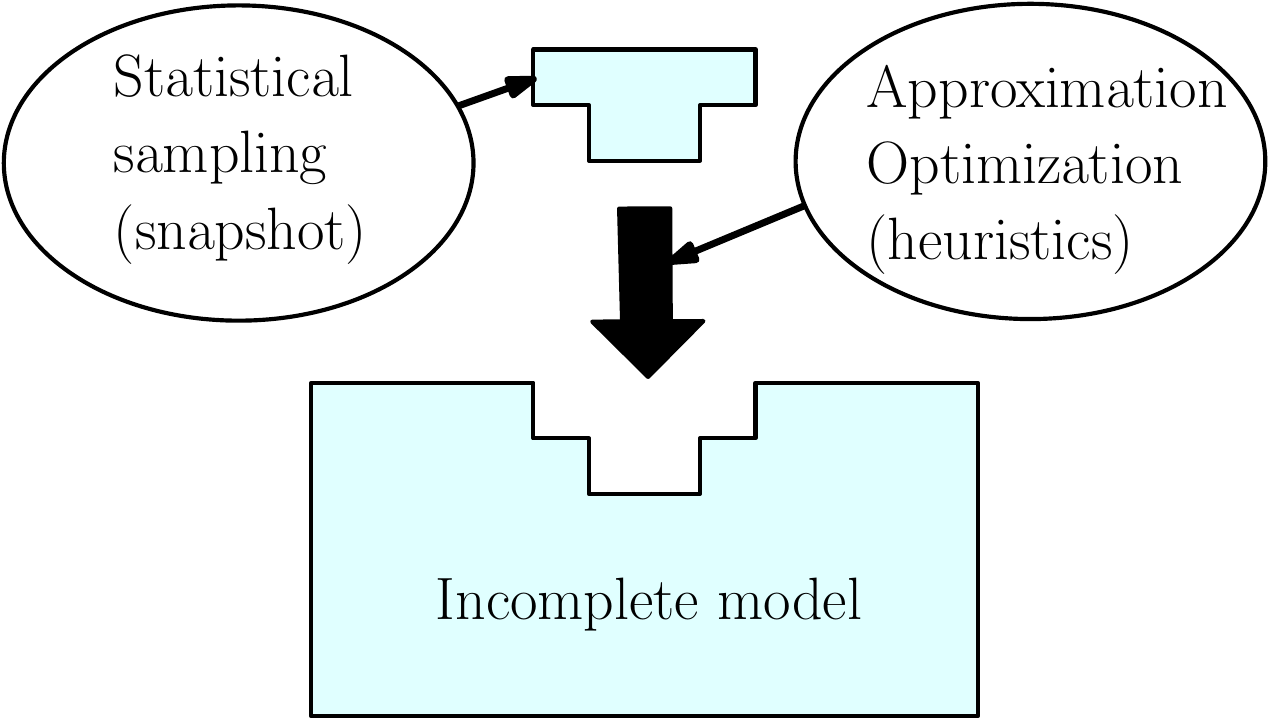}
 	 \caption{Schematic representation of the model completion simheuristic}
 	 \label{fig:schema}
 \end{center}	
 \end{figure}

Our \emph{simheuristic approach} is conceptualized in Figure \ref{fig:schema}. An incomplete \emph{dynamic} simulation model is executed. At the end of the execution, the missing part is replaced with some \emph{static} information, which is obtained by means of statistical sampling and subsequent numerical processing (i.e. optimization, approximation, etc.). The role of heuristics here is to provide the missing data of the dynamic simulation model, by solving the associated approximation$/$optimization problem. 

The use of heuristics in this case is justified for two reasons: 
\begin{enumerate}
\item The approximation$/$optimization problem may be computationally intractable (such as \Rep).
\setlength{\itemsep}{0pt}
\item Even if the problem were tractable, the parameters to be approximated are only \emph{estimates} of the real parameters, hence it is not necessary to find an exact solution; it suffices to find an approximate solution within some approximation threshold. Therefore we can opt for the most efficient solution method, which in our case is a combination of simple heuristics.   
\end{enumerate}

It would be interesting to adapt this scheme to other situations in order to better assess its usefulness and generality. For example, a straightforward extension of this work would be to verify how this approach fits other social networks equipped with the endorsement option, such as \ResearchGate. 

In our particular case, there are several open questions related to the endorsement approximation problem. First of all, it would be interesting to establish further complexity properties of the problem, as well as its relationship with other combinatorial optimization problems. We have been unable to find any reference to \Rip \ in the literature after 1990, which leads us to suspect that very little is known about its complexity, besides the fact that it is $\NP$-complete. Additionally, the scope and effectiveness of our heuristic solution should be determined more precisely. Finally, other heuristics may be explored, which might yield better results. 





\section*{Appendix: Some definitions and notation}
\label{app:def}

We review here the main definitions and notational conventions related to graph theory and complex networks. 

A \emph{directed graph}, or {\em digraph\/} $D=(V,A)$ is a finite nonempty set $V$ of objects called {\it vertices\/} and a set $A$ of ordered pairs of vertices called {\it arcs\/}. The {\it order\/} of $D$ is the cardinality of its set of vertices $V$. If $(u,v)$ is an arc, it is said that $u$ is {\em adjacent to\/} $v$, and $v$ is {\em adjacent from\/} $u$. The set of vertices that are adjacent from a given vertex $u$ is called the {\em out-neighbourhood\/} of $u$. It is denoted by $N^{+}(u)$ and its cardinality is the {\em out-degree\/} of $u$, $d^{+}(u)$. The {\em in-neighbourhood\/} of $v$ (denoted $N^{-}(u)$) is defined in an obvious way. 

Given a digraph $D=(V,A)$ of order $n$, the adjacency matrix of $D$ is an $n \times n$ matrix $\mathbf{M}=(m_{ij})_{n \times n}$ with $m_{ij}=1$ if $(v_i,v_j) \in A$, and $m_{ij}=0$ otherwise. The sum of all elements in the $i$-th row of $M$ will be denoted $\Sigma m_{i*}$, and it corresponds to $d^{+}(v_i)$. 


The distance from vertex $u$ to vertex $v$ is the length of the shortest path from $u$ to $v$. The diameter of $D$ is the maximum distance among all ordered pairs of vertices $(u,v)$. 

An \emph{undirected graph} (or simply a \emph{graph}) $G=(V,E)$ is a finite nonempty set $V$ of objects called {\it vertices\/} and a set $E$ of unordered pairs of vertices called {\it edges\/}. Again, the {\it order\/} of $G$ is the cardinality of its set of vertices $V$. If $(u,v)$ is an edge, we say that $u$ and $v$ {\em adjacent\/} to each other. The set of vertices that are adjacent to a given vertex $u$ is called the {\em neighbourhood\/} of $u$. It is denoted by $N(u)$, and its cardinality is the {\em degree\/} of $u$, $d(u)$.

A graph $G=(V,E)$ can be viewed as a \emph{symmetric} digraph, i.e. a digraph $D=(V,A)$ where for each arc $(u,v) \in A$, the reverse arc $(v,u)$ is also contained in $A$. With this view, the definitions of adjacency matrix, distance and diameter carry over naturally to graphs. The reader is referred to \cite{CL} for additional concepts on graphs and digraphs. 

Graphs and digraphs are important as models of complex real-life networks, such as social, biological, or communication networks. In those settings, vertices are usually called \emph{nodes} and arcs or edges are usually called \emph{links}. Although complex networks are conceptually graphs (or digraphs), and as such they inherit all graph properties (e.g. diameter), the study of complex networks usually requires additional mathematical tools and techniques, such as the tools of probability theory and statistics, which are less frequent in graph theory. 

For example, in a graph of manageable size we would be interested in determining the exact \emph{degree sequence}, i.e. the sequence of the degrees of all vertices. On the other hand, in a large complex network we would be more interested in describing the set of degrees as a probability distribution, such as a \emph{power law} $\mathbb{P}[k] \sim k^{-\lambda}$, where $k$ represents the degree of the nodes, and $\lambda$ is a constant parameter. 

Power law distributions are usually called \emph{scale-free} because of the relation $(ak)^{-\lambda} = a^{-\lambda}k^{-\lambda} \sim k^{-\lambda}$. This type of distribution is typical of networks generated by \emph{preferential attachment}: When a new node arrives, the probability of attaching it to an existing node $x$ is proportional to the degree of $x$. In other words, the rich get richer with time. For other degree distributions, other models of generation, and additional concepts related to complex networks, see \citet{CH}. 

\end{document}